\documentclass[letterpaper,twocolumn,10pt]{article}
\usepackage{usenix}
\usepackage{hyperref}
\usepackage{tikz}
\usepackage{xspace}
\usepackage{fontawesome5}
\usepackage{amsmath}
\usepackage{amssymb}
\usepackage{subcaption}
\usepackage[table]{xcolor}
\usepackage{booktabs}
\usepackage{enumitem}
\usepackage[ruled,linesnumbered]{algorithm2e}
\usepackage{pgfplots}
\usepgfplotslibrary{groupplots}
\usepackage{listings}
\usepackage{duckuments}
\usepackage{graphicx}
\usepackage{multirow}
\usepackage{fancyvrb}
\lstdefinelanguage{json}{
    morestring=[b]",
    morecomment=[l]{//},
    literate=
        *{:}{{{\color{black}{:}}}}{1}
         {,}{{{\color{black}{,}}}}{1}
}
\pgfplotsset{compat=1.18}
\usetikzlibrary{calc,positioning,backgrounds,fit,shapes.geometric}

\DeclareMathOperator*{\argmin}{arg\,min}

\newcommand{\myFaIcon}[1]{\makebox[1.75em][c]{\faIcon[regular]{#1}}}
\tikzset{myarrow/.style={->, thick}}
\newcommand{\baseTODO}[3]{\bgroup\color{#2}[\textsc{#1:} #3]\egroup\xspace}
\newcommand{\plainTODO}[2]{\bgroup\color{#1}[#2]\egroup\xspace}

\newcommand{\aim}{\text{\textsc{AIM}}\xspace}
\newcommand{\prisma}{\text{\textsc{Prisma}}\xspace}

\newcommand{\mwempgm}{\text{\textsc{MWEM+PGM}}\xspace}

\newif\ifdraft
\drafttrue
\begin{document}
\date{}

\title{\Large \bf PRISMA: Differentially Private Synthetic In-Car
Telemetry\\via Differentiable Simulation}

\author{
{\rm Kilian Zeiseweis$^*$}\\
BIH at Charité \\
kilian.zeiseweis@bih-charite.de
\and
{\rm Nele Sophie Brügge$^*$}\\
DFKI \\ %
nele.bruegge@dfki.de
\and
{\rm Marven Kummerfeldt} \\
University of Luebeck \\
marven.kummerfeldt@uni-luebeck.de
\and
{\rm Ali Raza} \\
HRI Europe GmbH \\
ali.raza@honda-ri.de
\and
{\rm Esfandiar Mohammadi} \\
University of Luebeck \\
esfandiar.mohammadi@uni-luebeck.de
}

\maketitle

\begin{abstract}
Differentially private synthesis of in-car energy consumption data is challenging because the output is high-dimensional, and known lower bounds show that the privacy-utility trade-off degrades with the square root of the number of output dimensions. However, we observe that the intrinsic dimensionality of in-car energy consumption is much lower than that of the raw trace: given a small set of contextual personal variables--such as start point, end point, the car model, and the driving behavior--the consumption profile can be largely explained by publicly available map data (including elevation) combined with a physical model of vehicle energy use. We exploit this low intrinsic dimensionality in PRISMA, a Privacy-preserving In-car Synthesis and Modelling Algorithm. PRISMA involves three stages: 
(i) energy consumption traces are encoded into a low-dimensional latent space; 
(ii) the distribution of latent vectors is synthesized using the differentially private tabular data synthesis algorithm AIM, and (iii) samples drawn from this distribution are decoded back into full energy consumption traces via a differentiable physics-based simulation, a step that consumes no additional privacy budget since it uses only public information. By using a compact, physically-grounded latent representation of the traces, PRISMA significantly reduces the effective output dimension. Our experimental results show that PRISMA achieves for a competetive $\varepsilon = 0.5$ a better 
data synthesis performance than prior tabular and marginal-based DP synthesis baselines by preserving the diversity of the original power consumption profiles without collapsing to a small set of modes.
\end{abstract}

\section{Introduction}

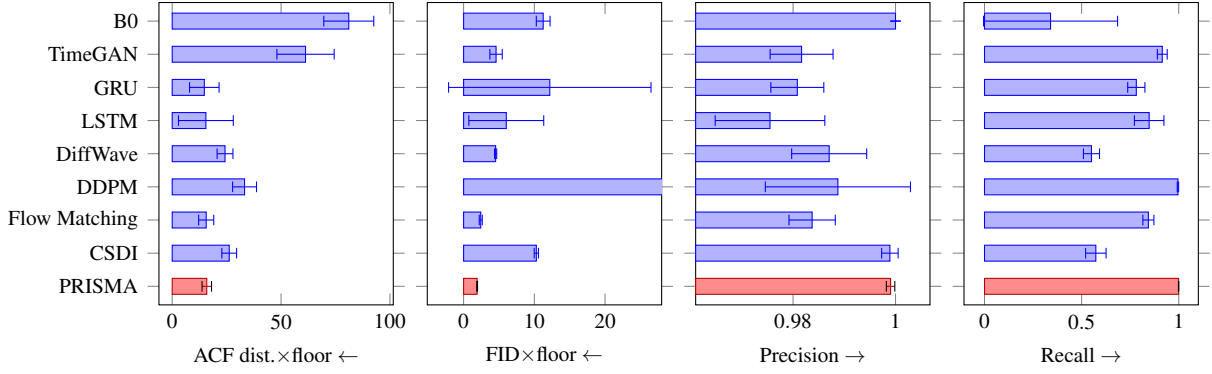
\begin{figure*}
    \centering
\begin{tikzpicture}
\begin{groupplot}[
    group style={group size=4 by 1, y descriptions at=edge left,
                 horizontal sep=0.45cm},
    xbar,
    symbolic y coords={{B0},{TimeGAN},{GRU},{LSTM},{DiffWave},{DDPM},{Flow Matching},{CSDI},{PRISMA}},
    ytick={{B0},{TimeGAN},{GRU},{LSTM},{DiffWave},{DDPM},{Flow Matching},{CSDI},{PRISMA}}, y dir=reverse, enlarge y limits=0.07,
    width=3.1cm, height=4cm, scale only axis,
    tick label style={font=\footnotesize},
    label style={font=\footnotesize},
]
\nextgroupplot[xlabel={ACF dist.$\times$floor $\leftarrow$}, bar width=6pt,
               error bars/x dir=both, error bars/x explicit]
\addplot+[bar shift=0pt, error bars/x dir=both, error bars/x explicit] coordinates {
    (81.099,{B0}) +- (11.472,0)
    (61.2401,{TimeGAN}) +- (13.1523,0)
    (14.7457,{GRU}) +- (6.78203,0)
    (15.4696,{LSTM}) +- (12.5883,0)
    (24.2592,{DiffWave}) +- (3.63983,0)
    (33.2304,{DDPM}) +- (5.48671,0)
    (15.6156,{Flow Matching}) +- (3.46658,0)
    (26.1912,{CSDI}) +- (3.39405,0)
};
\addplot[bar shift=0pt, draw=red!70!black, fill=red!45, error bars/x dir=both, error bars/x explicit] coordinates {
    (15.8837,{PRISMA}) +- (2.17654,0)
};
\nextgroupplot[xlabel={FID$\times$floor $\leftarrow$}, bar width=6pt, xmax=28,
               error bars/x dir=both, error bars/x explicit]
\addplot+[bar shift=0pt, error bars/x dir=both, error bars/x explicit] coordinates {
    (11.2517,{B0}) +- (0.971258,0)
    (4.58926,{TimeGAN}) +- (0.873185,0)
    (12.1887,{GRU}) +- (14.3091,0)
    (6.037,{LSTM}) +- (5.29237,0)
    (4.52719,{DiffWave}) +- (0.152473,0)
    (850.675,{DDPM}) +- (321.032,0)
    (2.42577,{Flow Matching}) +- (0.211757,0)
    (10.282,{CSDI}) +- (0.300877,0)
};
\addplot[bar shift=0pt, draw=red!70!black, fill=red!45, error bars/x dir=both, error bars/x explicit] coordinates {
    (1.91626,{PRISMA}) +- (0.0557562,0)
};
\nextgroupplot[xlabel={Precision $\rightarrow$}, bar width=6pt,
               error bars/x dir=both, error bars/x explicit]
\addplot+[bar shift=0pt, error bars/x dir=both, error bars/x explicit] coordinates {
    (1,{B0}) +- (0,0)
    (0.981667,{TimeGAN}) +- (0.00614184,0)
    (0.980833,{GRU}) +- (0.00516935,0)
    (0.9755,{LSTM}) +- (0.0107005,0)
    (0.98705,{DiffWave}) +- (0.00732103,0)
    (0.98875,{DDPM}) +- (0.014162,0)
    (0.983725,{Flow Matching}) +- (0.00451795,0)
    (0.9989,{CSDI}) +- (0.00160935,0)
};
\addplot[bar shift=0pt, draw=red!70!black, fill=red!45, error bars/x dir=both, error bars/x explicit] coordinates {
    (0.999033,{PRISMA}) +- (0.000825967,0)
};
\nextgroupplot[xlabel={Recall $\rightarrow$}, bar width=6pt,
               error bars/x dir=both, error bars/x explicit]
\addplot+[bar shift=0pt, error bars/x dir=both, error bars/x explicit] coordinates {
    (0.340025,{B0}) +- (0.345064,0)
    (0.915167,{TimeGAN}) +- (0.0253848,0)
    (0.7815,{GRU}) +- (0.0443076,0)
    (0.847667,{LSTM}) +- (0.0760278,0)
    (0.5506,{DiffWave}) +- (0.0413868,0)
    (0.9959,{DDPM}) +- (0.00367287,0)
    (0.84365,{Flow Matching}) +- (0.0283862,0)
    (0.57315,{CSDI}) +- (0.0526802,0)
};
\addplot[bar shift=0pt, draw=red!70!black, fill=red!45, error bars/x dir=both, error bars/x explicit] coordinates {
    (0.999367,{PRISMA}) +- (0.000865384,0)
};
\end{groupplot}
\end{tikzpicture}
    \caption{\textbf{Generation fidelity and diversity comparison under differential privacy ($\epsilon = 0.5$).} Evaluation of PRISMA against baseline synthesis models ($\leftarrow{}$~means lower is better and $\rightarrow{}$ higher is better) across the three \emph{fidelity} metrics ACF distance (autocorrelation, correlations over time), FID (Fréchet distance, Gaussian trajectory distribution discrepancy), and sample-level Precision (closeness to a real data points), and the \emph{diversity} metric Recall (coverage of real data distribution). Error bars denote standard deviation across 15 runs. ACF and FID are reported as multiples of the empirical null floor, where a value of $k\times\text{floor}$ indicates the distance is $k$ times worse than the distance between two splits of real holdout data.}
    \label{fig:barplot}
\end{figure*}

Modern vehicles continuously generate large volumes of telemetry data through onboard sensors and electronic control units. These data capture driving behavior, vehicle dynamics, environmental conditions, and energy consumption, supporting a wide range of applications: energy-efficiency modeling, intelligent transportation systems, and driver-assistance technologies. However, vehicle telemetry is inherently privacy-sensitive. Fine-grained speed profiles, location traces, and auxiliary sensor readings can reveal an individual's daily routine, home address, and behavioral patterns.  %
 
Synthetic data generation enables data sharing while mitigating privacy risks. By releasing artificial data that preserves important statistical properties of real-world observations, organizations can support downstream research without exposing sensitive records. 
A natural way to obtain formal privacy guarantees is to apply an established differentially private (DP) tabular synthesis algorithm directly to the raw trace. For time series data, this is unattractive: known DP lower bounds show that the achievable privacy--utility trade-off degrades with the square root of the number of released dimensions. Therefore, directly synthesizing every time step is a poor fit for high-resolution time series data.
Especially for more complex time series, like vehicle telemetry, generating synthetic data that is simultaneously realistic and accompanied by formal privacy guarantees remains an open challenge.

Our key observation is that an energy-consumption trace for a single trip has much lower intrinsic dimensionality than the trace itself. 
Given the trip's start and end coordinates, the vehicle model, and a small number of behavioral parameters (how aggressively the trip was driven and how much detouring occurred), a physics simulation over publicly available map data (road network, elevation) and a public vehicle catalog can reconstruct a trace that closely resembles the real one. 

We propose \prisma as an algorithm that synthesizes only the low-dimensional descriptor under DP and reconstructs the trace from this descriptor as a public, privacy-free post-processing step. 
Our algorithm works in three stages: 
In the \emph{encoding} stage, using public data, a differentiable physics simulation is fit to each real trace. The simulation produces a compact per-trip descriptor (start/end point, aggressiveness, detour factor, and a discrete vehicle choice) via gradient-based optimization. 
In the \emph{DP synthesis} stage, the resulting table of descriptors is used to train the query-based DP tabular synthesis algorithm AIM~\cite{aim}, selecting informative low-order marginal queries adaptively. A synthetic descriptor table is then sampled directly from the fitted model. 
In the \emph{decoding} stage, the same physics simulation is re-run forward on each synthetic descriptor to produce a full synthetic energy trace. 
Since this step only uses public inputs and the already private descriptor, it is DP-preserving post-processing and costs no additional privacy budget~\cite{dwork_algorithmic_2014}.

Prior DP synthesis methods target tabular data~\cite{privsyn,zhang_privbayes_2017,truda_generating_2023,aim} or time series~\cite{arnout_differentially_2021, esteban_real-valued_2017} directly. While the latter are tailored to high-dimensional temporal data, they do not exploit prior domain knowledge to reduce its dimensionality. Separately, simulation-based approaches are widely used to study driver behavior and vehicle energy consumption.
\prisma combines both and uses a differentiable physics simulation to encode each trace into a compact descriptor, which can then be synthesized with standard DP tabular methods. To our knowledge, combining simulation-based dimensionality reduction with DP synthesis has not yet been explored.

We make the following contributions:
\begin{itemize}
  \item We identify that in-car energy telemetry has low intrinsic dimensionality relative to the raw trace and show that this can be exploited by decoupling encoding/decoding (physics simulation) from distribution synthesis (tabular DP synthesis on a compact descriptor).
  \item We design \prisma's encoder, which fits a per-trip descriptor to a real trace via gradient-based optimization through a differentiable physics simulation, and its decoder, which reconstructs a synthetic trace from a descriptor using the same simulation as public post-processing.
  \item We evaluate \prisma against a range of suitable (DP) time series synthesis models. 
\end{itemize}

\section{Background}

\subsection{Differential Privacy}
Differential privacy (DP)~\cite{dwork2006, dwork_algorithmic_2014} provides a formal definition of privacy loss when releasing statistics computed over a sensitive dataset. 
A randomized mechanism $\mathcal{M}$ satisfies $(\varepsilon,\delta)$-DP if, for any two neighboring datasets $D$ and $D'$ differing in at most one individual's record and any output set $S$:
\[
  \Pr[\mathcal{M}(D) \in S] \;\leq\; e^{\varepsilon} \cdot \Pr[\mathcal{M}(D') \in S] + \delta.
\]
The parameter $\varepsilon$ bounds privacy loss (smaller is stronger) and $\delta$ is a failure probability, set to $10^{-5}$ throughout this work. The \emph{Gaussian mechanism} achieves $(\varepsilon,\delta)$-DP by adding noise $\mathcal{N}(0, \Delta^2 \sigma^2)$ calibrated to the $\ell_2$-sensitivity $\Delta$ of the released function~\cite{dwork_algorithmic_2014}. 
A key property used in \prisma is \emph{post-processing}: any computation applied to the output of a DP mechanism without further access to the sensitive data is itself DP at the same $(\varepsilon,\delta)$~\cite{dwork_algorithmic_2014}.
This is what allows \prisma's decoding stage to run without spending any additional privacy budget.

\subsection{Differentiable Programming}
\label{sec:background-diffprog}
Differentiable programming enables gradients to be computed automatically through an entire program, including programs with control flow and data structures~\cite{blondel_elements_2025}. The parameters of a simulation can therefore be adjusted with gradient-based optimization so that its output better matches observed data.
In the context of this work, gradients of the simulated trace with respect to the simulation parameters are used to fit a low-dimensional representation of each trip. A practical challenge is posed by operations that are not differentiable, such as conditions, comparisons, and discrete switches. Before gradient-based optimization can be applied, these operations are replaced with smooth, differentiable approximations.

\paragraph{Differentiable simulation.} A differentiable simulation can be represented as a discrete-time state-space model~\cite{wang_state_2025}. At time step $k$, the current state $x(k)$ and input $u(k)$ are mapped to the next state by $x(k+1) := f(x(k), u(k))$ and to the output by $y(k+1) := g(x(k), u(k))$, where $f$ is the state-transition function and $g$ is the output function. The simulation $S := (f,g)$ is differentiable when both mappings are differentiable with respect to their inputs and parameters.  If conditions, comparisons, or discrete switches are required internally, they are replaced with smooth relaxations, for example, using sigmoid functions for binary conditions or softmax for discrete choices, so that gradients can be propagated through the entire simulation.

\section{Related Work}
 
\subsection{Synthetic Time Series Generation} 

\paragraph{Non-Private time series generation.}  %
Generative modeling of time series data has developed along architectures that explicitly account for temporal dependencies. Early neural approaches mainly used recurrent~\cite{hochreiter_long_1997, cho_learning_2014} and adversarial architectures~\cite{yoon_time-series_2019, lin_using_2020}. TimeGAN~\cite{yoon_time-series_2019} combines an autoencoder-based embedding with supervised and adversarial objectives to jointly learn the latent representation and temporal dynamics of sequences, while TimeVAE~\cite{desai_timevae_2021} instead uses a variational autoencoder.
DoppelGANger~\cite{lin_using_2020} models static attributes together with time-varying signals using a generative adversarial architecture with autoregressive generation.

More recently, diffusion models have emerged as a powerful alternative. DiffWave~\cite{kong_diffwave_2021} applies denoising diffusion to one-dimensional temporal signals, while CSDI~\cite{tashiro2021csdi} uses temporal and feature-wise attention to model multivariate series. Diffusion-TS~\cite{yuan_diffusion-ts_2024} combines diffusion with a decomposition into trend- and Fourier-based seasonal components, and DiffTime~\cite{coletta_constrained_2023} conditions the diffusion process on temporal constraints such as trends and fixed points. Flow-matching approaches use a related formulation based on learning continuous transport paths between distributions~\cite{liu_flow_2023,lipman_flow_2023}. 

\paragraph{Differentially-private approaches.}

Differentially private synthetic data generation can be broadly categorized into query-based and model-based methods. 
Query-based methods release input data statistics and construct a synthetic dataset consistent with them. 
Model-based methods learn a generative model and sample from it~\cite{pgm,zhang_privbayes_2017,pategan,dpgan}. 
Query-based approaches~\cite{rap, rapPlus, aim} commonly use marginal queries as they provide an expressive representation of attribute distributions while decomposing synthesis into tractable subproblems. Prior studies indicate that marginal-based methods often achieve a strong privacy-utility trade-off for tabular data~\cite{tabBench}. 

A prominent class of \emph{query-based methods} combines marginal selection with probabilistic graphical models. \mwempgm~\cite{pgm} constructs a junction tree from a workload of noisy marginals and generates synthetic data from the resulting graphical model. It iteratively identifies poorly approximated marginals, measures them privately, and refines the graphical model. \aim~\cite{aim} improves this framework through informed initialization, workload-aware candidate selection, adaptive budget allocation, and an improved privacy analysis.

\emph{Model-based methods} learn a generative model under differential privacy and sample from it. DPGAN~\cite{dpgan} applies DP-SGD to the discriminator, clipping per-example gradients and adding Gaussian noise. PATE-GAN~\cite{pategan} replaces the discriminator with a PATE ensemble of teachers whose noisy aggregated labels train a student discriminator. For tabular data, CTGAN~\cite{ctgan} uses conditional generation and a mode-specific loss to handle mixed discrete–continuous columns. Its DP variants, DP-CTGAN and PATE-CTGAN, combine this architecture with DP-SGD or PATE, respectively~\cite{fang_dpctgan}.
Recent DP time series models follow the same DP-SGD approach. For example, \cite{esteban_real-valued_2017,arnout_differentially_2021} apply DP-SGD to recurrent and GAN-based time series generators.

\subsection{Simulation-Based Approaches for Driver and Vehicle Modeling}
Simulation-based inference has received growing attention in recent years, particularly for inferring the behaviour of physical systems from observed data~\cite{cranmer_frontier_2020, hao_physics-informed_2023}.

Differentiable programming frameworks~\cite{blondel_elements_2025} have recently made it practical to differentiate through such simulations end-to-end by replacing non-smooth operations (routing decisions, discrete choices) with smooth or straight-through approximations.
\prisma applies these techniques to fit a compact descriptor to each real trip and to reconstruct traces from synthetic descriptors, and combines this with DP tabular synthesis on the descriptor. Prior work on privacy in vehicular settings has largely focused on location privacy~\cite{Yi2025} and anonymization of GPS trajectories \cite{LIU2021115120}. Combining a physics-based reduction of trace dimensionality with DP tabular synthesis, as \prisma does, has not been explored previously to our knowledge.

\section{PRISMA}
\label{sec:prisma}

\begin{figure*}[htbp]
\centering
\footnotesize
\begin{tikzpicture}[
  every node/.style={font=\footnotesize},
  stage/.style={draw, rounded corners=3pt, align=center, inner sep=3pt,
                minimum height=1.15cm, text width=1.85cm},
  data/.style={draw, rounded corners=3pt, align=center, inner sep=3pt,
               minimum height=1.15cm, text width=1.45cm, fill=gray!8},
  node distance=5mm and 5mm,
]
  \node[data] (real) {Real traces\\ $N$ trips};
  \node[stage, fill=red!12,    right=of real] (enc) {\textbf{Encode}\\ fit $3$ numbers per trip};
  \node[data,                  right=of enc]  (tab) {Descriptor table $N\!\times\!7$};
  \node[stage, fill=yellow!30, right=of tab]  (aim) {\textbf{Synthesize}\\ AIM on public bins};
  \node[data,                  right=of aim]  (syn) {Synthetic table $N\!\times\!7$};
  \node[stage, fill=green!18,  right=of syn]  (dec) {\textbf{Decode}\\ route \\$\to$ profile $\to$ driver $\to$ power};
  \node[data,                  right=of dec]  (out) {Synthetic traces};

  \draw[myarrow] (real) -- (enc);
  \draw[myarrow] (enc)  -- (tab);
  \draw[myarrow] (tab)  -- (aim);
  \draw[myarrow] (aim)  -- (syn);
  \draw[myarrow] (syn)  -- (dec);
  \draw[myarrow] (dec)  -- (out);

  \node[draw, dashed, rounded corners=3pt, fill=blue!5, align=center,
        text width=15.2cm, inner sep=4pt]
        (pub) at ($(real)!0.5!(out) + (0,1.85)$)
        {\textbf{Public, data-independent inputs} -- no privacy cost:
         OpenStreetMap drive graph over a fixed bounding box $\cdot$ EPA vehicle
         catalogue $\cdot$ vehicle power equation $\cdot$ efficiency $\eta = 0.30$
         $\cdot$ bin grids and ranges $\cdot$ the $500$\,s horizon $\cdot$ the row
         count $N$};
  \draw[myarrow, dashed] (pub.south -| enc.north) -- (enc.north);
  \draw[myarrow, dashed] (pub.south -| aim.north) -- (aim.north);
  \draw[myarrow, dashed] (pub.south -| dec.north) -- (dec.north);

  \begin{pgfonlayer}{background}
    \node[draw, dashed, rounded corners=4pt, fill=green!5, inner sep=5pt,
          fit=(syn)(dec)(out),
          label={[font=\scriptsize, anchor=north]south:post-processing: no privacy cost}]
          (post) {};
  \end{pgfonlayer}
\end{tikzpicture}
\caption{The \prisma pipeline. The encoder fits a low-dimensional representation of each real trace using a differentiable physics simulation. The resulting descriptor table is synthesized under differential privacy using AIM. Due to the post-processing property of DP, the decoder can reconstruct full synthetic trajectories from the synthetic representation without spending additional privacy budget.}
\label{fig:pipeline}
\end{figure*}
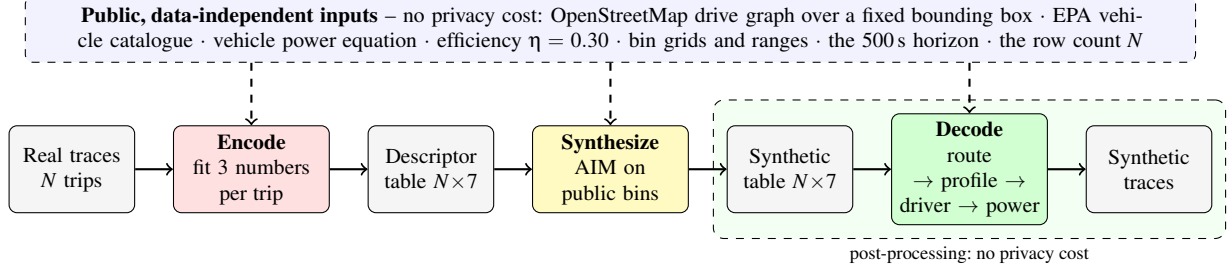

\prisma is designed to synthesize realistic in-car energy consumption traces. The main idea is to find a low-dimensional representation of the power trajectories that can be synthesized under differential privacy, and then to reconstruct the full traces from this representation using the differentiable simulation. Figure \ref{fig:pipeline} illustrates the overall pipeline. The full process of synthesizing a dataset of traces can be seen in three stages: \textit{(i) encoding}, \textit{(ii) synthesis}, and \textit{(iii) decoding}, which are described in detail in the following sections.

In general, \prisma can be seen as an autoencoder-like architecture, where the encoder maps a high-dimensional input (the real trace) to a low-dimensional latent space (the representation), and the decoder reconstructs the high-dimensional output (the synthetic trajectory) from the latent representation. The key difference is that the encoder and decoder are based on a differentiable physics simulation. This allows us to incorporate fine-grained domain knowledge about the underlying physical processes, which enables a realistic reconstruction of the traces from a low-dimensional representation that can be synthesised under differential privacy.

\subsection{Public Parameters}\label{sec:public-parameters}

One of the key ideas of \prisma is to leverage public domain knowledge about the physics of vehicles and the environment in which they operate to reduce the dimensionality of the representation. In order to reconstruct a realistic power trace from the low-dimensional representation of a trip, the forward simulation $\mathcal{S}$ requires a set of public parameters $\mathbf{p}$. In detail, the following public parameters are assumed to be available and are therefore not part of the private representation $\mathbf{d}$:

\begin{enumerate}[label=(\roman*)]
    \item A public road network. This is extracted from OpenStreetMap \cite{openstreetmap} (OSM) and contains the road segments, their lengths, and speed limits. It was processed using OSMnx \cite{boeing_osmnx_2025}. The road network is used to compute the route between the start and end points of the trip.
    \item A public elevation map. It is used to compute the slope of the road segments, which is necessary to compute the gravitational force acting on the vehicle. The elevation map was obtained from the Terrain Tiles archive of public elevation data \cite{terraintiles}.
    \item A public pool of candidate vehicles $\mathcal{V}$ derived from the EPA's fuel economy database \cite{epa_testcar}. It contains the mass, drag coefficient, and rolling resistance of a set of vehicles built and tested in the years 2000 to 2016. The idea behind using a public pool of vehicles is to reduce the dimensionality of the representation by only allowing a discrete set of vehicles to be chosen from, rather than allowing a continuous range of vehicle parameters. Further, they pin down the physics of the vehicle to realistic values, which is necessary to reconstruct a realistic power trace in the subsequent modeling steps.
\end{enumerate}

Together, they form a set of public parameters $\mathcal{P}$ that are used in the forward simulation $\mathcal{S}$ to compute the power trace from the low-dimensional representation of a trip. The public parameters are assumed to be known and do not contain any sensitive information about the individual trips, and therefore do not need to be protected under differential privacy.

\subsection{Latent Representation}
\label{sec:Latent Representation}

The latent representation, in contrast to the public parameters, is a low-dimensional representation of a trip that contains all the information necessary to reconstruct the power trace using the forward simulation $\mathcal{S}$. It is the result of the encoding stage (see Section~\ref{sec:encoding}) and is used as input to the decoding stage (see Section~\ref{sec:decoding}). The latent representation is designed to be low-dimensional, so that it can be synthesized under differential privacy without losing too much utility.

Each trip is summarised by a representation $\mathbf{d} = (\phi_s, \psi_s, \phi_e, \psi_e, \alpha, \delta, \kappa)$ with seven fields: the start and end latitude/longitude $(\phi, \psi)$, a scalar \emph{aggressiveness} $\alpha \in [0,1]$ describing how aggressively the trip was driven, a scalar \emph{detour} factor $\delta$ describing how much longer than the shortest path the route ran, and a categorical \emph{vehicle} choice $\kappa$ drawn from a released pool of EPA candidate vehicles $\mathcal{V}$.

Table~\ref{tab:descriptor} lists the fixed, public discretisation grids used when the descriptor table is later handed to AIM (Section~\ref{sec:aim}). These grids are chosen before any data is read and do not depend on the dataset, so choosing them costs no privacy budget.
\begin{table}[t]
\centering
\caption{Representative fields of domain $\mathcal{D}$ and their fixed public discretisation grids. The evaluated bounding box gives roughly 1\,km resolution per bin over the region covered by the ground-truth data.}
\label{tab:descriptor}
\small
\setlength{\tabcolsep}{3pt}
\begin{tabular}{llll}
\toprule
\textbf{Symbol} & \textbf{Field} & \textbf{Domain} & \textbf{Bins} \\
\midrule
$(\phi_s, \psi_s)$ & Start lat.\ / lon. & bbox\footnotemark & $32\times32$ \\
$(\phi_e, \psi_e)$ & End lat.\ / lon.   & bbox\footnotemark[\value{footnote}] & $32\times32$ \\
$\alpha$           & Aggressiveness     & $[0, 1]$              & 16 \\
$\delta$           & Detour             & $[1.0, 2.5]$          & 16 \\
$\kappa$           & Vehicle            & EPA pool              & $|\mathcal{V}|$ \\
\bottomrule
\end{tabular}%

\end{table}

\subsection{Forward Simulation Model}\label{sec:forward-model}

The physics of vehicles is well understood and can be expressed as a set of formulas and mechanistic models. The forward simulation $\mathcal{S}$ takes as input a representation of a trip $\mathbf{d} \in \mathcal{D}$ and a set of public parameters $\mathbf{p} \in \mathcal{P}$, and outputs a power trace $\mathbf{W} \in \mathbb{R}^{T}$, where $T$ is the number of time steps in the trajectory $\mathcal{S} : \mathcal{D} \times \mathcal{P} \;\longrightarrow\; \mathbb{R}^{T}$.

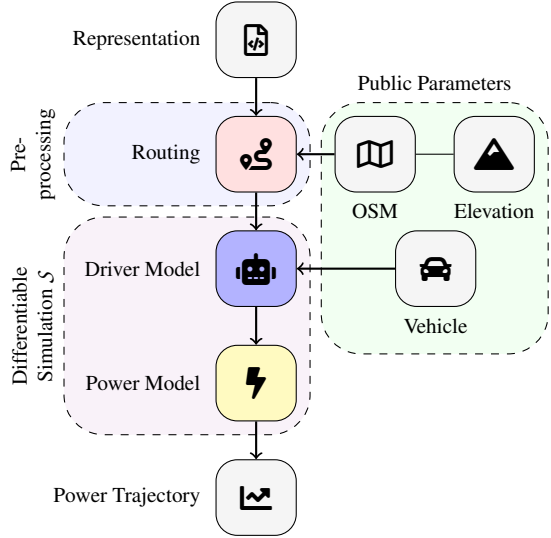
\begin{figure}[bt]
\centering
\footnotesize
\begin{tikzpicture}[
  stage/.style={draw, rounded corners=8pt, align=center, inner sep=3pt,
                minimum height=1cm, text width=0.85cm, font=\large},
  data/.style={draw, rounded corners=8pt, align=center, inner sep=3pt,
               minimum height=1cm, text width=0.85cm, fill=gray!8, font=\large},
  label/.style={align=center},
  node distance=5mm and 5mm,
]

\node[data] (repr) {\myFaIcon{file-code}};
\node[stage, fill=red!12, below=of repr] (routing) {\myFaIcon{route}};
\node[stage, fill=blue!30, below=of routing] (driver) {\myFaIcon{robot}};
\node[stage, fill=yellow!30, below=of driver] (power) {\myFaIcon{bolt}};
\node[data, below=of power] (trace) {\myFaIcon{chart-line}};

\node[data, right=0.5cmof routing] (osm) {\myFaIcon{map}};
\node[data, right=of osm] (elev) {\myFaIcon{mountain}};
\node[data, right=1.3cmof driver] (veh) {\myFaIcon{car}};

\node[label, left=0.1cmof repr] (repr-label) {Representation};
\node[label, left=0.1cmof routing] (routing-label) {Routing};
\node[label, left=0.1cmof driver] (driver-label) {Driver Model};
\node[label, left=0.1cmof power] (power-label) {Power Model};
\node[label, left=0.1cmof trace] (trace-label) {Power Trajectory};

\node[label, below=0.05of osm] (osm-label) {OSM};
\node[label, below=0.05of elev] (elev-label) {Elevation};
\node[label, below=0.05of veh] (veh-label) {Vehicle};

\begin{pgfonlayer}{background}
    \node[
        draw, 
        dashed,
        rounded corners=15pt, 
        align=center, 
        inner xsep=5pt, 
        inner ysep=5pt, 
        fit=(osm) (elev) (veh) (veh-label),
        label={[anchor=north east, scale=1.5, align=right]north east:Simulation },
        fill=green!5
    ] (public-data) {};
\end{pgfonlayer}

\begin{pgfonlayer}{background}
    \node[
        draw, 
        dashed,
        rounded corners=15pt, 
        align=center,
        inner xsep=5pt,
        inner ysep=5pt, 
        fit=(driver) (power) (driver-label) (power-label),
        fill=violet!5
    ] (simulation) {};
\end{pgfonlayer}

\coordinate (routing-align) at (driver-label.west |- routing.center);
\begin{pgfonlayer}{background}
    \node[
        draw,   
        dashed,
        rounded corners=15pt, 
        align=center, 
        inner xsep=5pt,
        inner ysep=5pt, 
        fit=(routing) (routing-label) (routing-align),
        fill=blue!5
    ] (pre-processing) {};
\end{pgfonlayer}

\node[label, above=0.05cm of public-data] (public-data-label) {Public Parameters};
\node[label, left=0.4of simulation, rotate=90, align=center, anchor=center] (simulation-label) {Differentiable \\ Simulation $\mathcal{S}$};
\node[label, left=0.4of pre-processing, rotate=90, align=center, anchor=center] (pre-processing-label) {Pre-\\processing};

\draw[myarrow] (repr) -- (routing);
\draw[myarrow] (routing) -- (driver);
\draw[myarrow] (driver) -- (power);
\draw[myarrow] (power) -- (trace);

\draw[myarrow] (osm) -- (routing);
\draw[-] (elev) -- (osm);
\draw[myarrow] (veh) -- (driver);

\end{tikzpicture}
\caption{The forward simulation $\mathcal{S}$ takes as input a trip representation $\mathbf{d}$ and public parameters $\mathbf{p}$ to produce a power trace. The pre-processing step computes the route from the start and end points of the trip, which is then used by the differentiable driver and vehicle models to compute the power trace.}
\label{fig:forward-simulation}
\end{figure}

Figure \ref{fig:forward-simulation} illustrates the simulation process. The following five steps are performed in order to compute the power trace from the trip representation $\mathbf{d}$ and public parameters $\mathbf{p}$.

\paragraph{1. Routing.} \label{sec:routing} The routing is part of the pre-processing step before the actual simulation of the vehicle. The idea is to take the start and end points of the trip and compute the shortest path between them using the public road network. This route is then used in the differentiable simulation to compute the power trace. The algorithm works as follows:

\begin{enumerate}[label=(\roman*)]\label{alg:routing}
  \item \textbf{Snap start and end points to the road network.} The start and end coordinates $(\phi_s, \psi_s)$ and $(\phi_e, \psi_e)$ of the trip are snapped to the nearest road segments in the public road network. This ensures that the routing is performed on valid roads.
  \item \textbf{Route computation.} Using the snapped start and end points, the shortest path is computed using Dijkstra's algorithm for weighted graphs, where the weights are based on the length of the road segments. The resulting route is a sequence of road segments that connect the start and end points.
  \item \textbf{Unfold the route.} The resulting route is unfolded into a sequence of waypoints along the road segments, resulting in a sequence of edges and their corresponding arc lengths $l_i$. The total length of the route is then computed as $L = \sum_i l_i$. From this, a sequence of arc lengths $\mathbf{s} = (s_1, \dots, s_S)$ is constructed, where $S$ is the number of waypoints along the route and $s_i = \sum_{j=1}^{i} l_j$.
  \item \textbf{Sample Speed Limit and Elevation.} For each edge in the route, the speed limit $\mathbf{u}$ and elevation $\boldsymbol{\theta}$ are sampled from the public road network data. This information is used to compute the drivable speed and the energy consumption in the subsequent steps.
  \item \textbf{Constrain the drivable speed}. Each junction along the route that has a traffic signal or stop sign lowers the drivable speed $\mathbf{u}$ to a creep speed $u_{\mathrm{stop}}$ for a short distance. The drivable speed is computed as a smooth envelope that respects the speed limits and the junction constraints.
\end{enumerate}

This information is then used by the differentiable driver and the differentiable vehicle model to compute the power trajectory.

\paragraph{2. Modeling the driver.} As shown in Figure \ref{fig:forward-simulation}, the second step of the forward simulation and the first step that is part of the differentiable simulation is the driver model. It consumes the route information $(\mathbf{s}, \mathbf{u}, \boldsymbol{\theta}, L)$ from the previous routing step and the aggressiveness $\alpha$, the detour factor $\delta$, and the vehicle index $\kappa$ from the trip representation $\mathbf{d}$ to compute a trace of the actual vehicle's speed and acceleration over time. 

The procedure for the driver model is shown in Algorithm \ref{alg:driver}, which was inspired by Gipps' car-following model~\cite{GIPPS1981105}. The driver model is a simple proportional controller that tries to follow the speed limit while respecting the vehicle's acceleration and deceleration limits. The aggressiveness $\alpha$ scales the maximum acceleration and deceleration, while the detour factor $\delta$ scales the length of the route. The driver model computes the target speed at each time step based on the current position along the route and the remaining distance to the end of the trip. The actual speed is then updated based on the target speed and the acceleration limits. 

\begin{algorithm}[t!]
\caption{\faRobot \ Driver Model}
\label{alg:driver}
\KwIn{Profile $(\mathbf{s}, \mathbf{u}, \boldsymbol{\theta}, L)$ of $S$ points from
      the Algorithm in Section~\ref{sec:routing}; detour $\delta$, aggressiveness $\alpha$, vehicle index $\kappa$; horizon $T$, time step $\Delta t$}
\KwData{Public constants: acceleration ceiling $A$, brake fraction $\beta$, control gain $K$}
\KwOut{Power trace $\hat{\mathbf{p}} \in \mathbb{R}^{T}$}
\BlankLine
$\mathbf{s} \leftarrow \delta\ \cdot \mathbf{s}$;\quad $L \leftarrow \delta \cdot L$;\quad
$\boldsymbol{\theta} \leftarrow \boldsymbol{\theta}/\delta$;\quad
$\Delta s \leftarrow L/(S-1)$
   \tcp*{Scale route for detour $\delta$}
$a_{\max} \leftarrow \alpha\,A$;\quad $b \leftarrow \beta\,a_{\max}$
   \tcp*{acceleration ceiling and comfortable braking}
$v^{\max}_i \leftarrow \min\!\big(u_i,\, \sqrt{(v^{\max}_{i+1})^2 + 2b\,\Delta s}\big)$
   for $i = S,\dots,1$, with $v^{\max}_{S+1} = 0$
   \tcp*{backward pass: brake into what lies ahead}
\BlankLine
$x \leftarrow 0$;\quad $v \leftarrow 0$
   \tcp*{position along the route (m) and speed (m/s)}
\For{$t \leftarrow 1$ \KwTo $T$}{
   $d \leftarrow \max(L - x - v\,\Delta t,\; 0)$
      \tcp*{distance still to drive, one step ahead}
   $v^{\star} \leftarrow \min\!\big(v^{\max}(x),\; \sqrt{2 b\, d}\big)$
      \tcp*{target: road ahead vs.\ arrival}
   $\alpha \leftarrow \mathrm{clip}\big(K\,(v^{\star} - v),\; -a_{\max},\; a_{\max}\big)$
      \tcp*{acceleration this step (m/s$^2$)}
   $\hat{p}_t \leftarrow vehiclePower(\kappa, v, \alpha, \theta(x))$
      \tcp*{mechanical power at the wheels (W)}
   $x \leftarrow x + v\,\Delta t$;\quad $v \leftarrow \max(v + \alpha\,\Delta t,\; 0)$\;
}
\Return $\hat{\mathbf{p}}$\;
\end{algorithm}

While the driver model is a simplification of real-world driving behavior, it captures the essential dynamics of acceleration and deceleration in response to speed limits and road conditions. The model is differentiable with respect to the aggressiveness $\alpha$ and the detour factor $\delta$, allowing for gradient-based optimization during the encoding stage.

\paragraph{3. Modelling the vehicle.}

During the forward simulation in Algorithm \ref{alg:driver}, the vehicle model computes the power drawn from the fuel tank or battery based on the current speed, acceleration, and road slope. Therefore, the vehicle model takes as input the vehicle index $\kappa$, the current speed $v$, the current acceleration $\alpha$, and the road slope $\theta$ to compute the power at time step $t$. The calculation is done in two steps:

\begin{enumerate}
  \item Compute the mechanical power at the wheels $P_{\mathrm{wheel}}$ based on the vehicle mass, frontal drag area, rolling-resistance coefficient, and the gravitational force due to the road slope using Eq.~\ref{eq:roadload}.
  \item Compute the power drawn from the fuel tank or battery $P_{\mathrm{tank}}$ based on the mechanical power at the wheels, the marginal tank-to-wheel efficiency $\eta$, the recuperated fraction $r$, and the idle power draw $P_{\mathrm{idle}}$ using Eq.~\ref{eq:tank}.
\end{enumerate}

\begin{equation}
  P_{\mathrm{wheel}} = \Big( m v + \tfrac{\rho}{2} C_d A\, v^2 + B v
      + C_{rr}\, m g + m g \sin\theta \Big)\, v .
  \label{eq:roadload}
\end{equation}

The reason behind this two-step calculation is that the mechanical power at the wheels is a function of the vehicle's physical properties and the driving conditions, while the power drawn from the fuel tank or battery also depends on the efficiency of the powertrain and other factors, like peripheral power draw. 

\begin{equation}
  P_{\mathrm{tank}} = \underbrace{\frac{\max(P_{\mathrm{wheel}}, 0)}{\eta}}_{\text{propulsive}}
      + \underbrace{r \min(P_{\mathrm{wheel}}, 0)}_{\text{recuperated}} + \underbrace{P_{\mathrm{idle}}}_{\text{idle}} .
  \label{eq:tank}
\end{equation}

The tractive force is computed using the standard longitudinal road-load form, which takes into account the vehicle's mass, frontal drag area, rolling-resistance coefficient, and the gravitational force due to the road slope. The tank power, on the other hand, is computed based on the mechanical power at the wheels, the marginal tank-to-wheel efficiency $\eta$, the recuperated fraction $r$ (in the case of hybrid or electric vehicles), and the idle power draw $P_{\mathrm{idle}}$. 

The vehicle model is differentiable with respect to the vehicle index $\kappa$, which is crucial for the overall differentiability of the forward simulation and enables the use of gradient-based optimization methods to fit the low-dimensional representation of the trips.

\subsection{Encoding}\label{sec:encoding}

During encoding, each observed power trace is mapped to a low-dimensional latent representation. A simulated trace is generated from this representation by the differentiable physics model, and its parameters are adjusted to minimize the discrepancy between the simulated and observed traces. Because the simulation is differentiable with respect to these parameters, this adjustment can be performed using gradient-based optimization.

\paragraph{Fitting the Representation.} \label{sec:fitting}

For each observed trip $i$, a descriptor is estimated to best reproduce the measured power trace $\mathbf{W}_i$. The trip endpoints $\mathbf{e}_i = (\phi_s, \psi_s, \phi_e, \psi_e)$ are read directly from the trace and held fixed. Only the remaining descriptor variables are fitted: aggressiveness $\alpha$, detour factor $\delta$, and candidate-vehicle index $\kappa$.

For a proposed set of values $(\alpha, \delta, \kappa)$, a simulated trace is produced by the forward simulation $\mathcal{S}$ from these variables, the fixed endpoints, and the public parameters $\mathbf{p}$. The fitted values are selected such that the loss $\mathcal{L}$ between the simulated and observed traces is minimized:

\begin{equation}
  (\alpha_i^{\star}, \delta_i^{\star}, \kappa_i^{\star}) =
  \argmin_{(\alpha, \delta, \kappa)}
  \mathcal{L}\big( \mathcal{S}(\mathbf{e}_i, \alpha, \delta, \kappa;\, \mathbf{p})\;
                   \mathbf{W}_i \big)
  \label{eq:optimization}
\end{equation}

The loss is defined in the next paragraph. This problem is solved independently for each trip. Because the simulator is differentiable with respect to the fitted variables, the reconstruction loss can be reduced iteratively by a gradient-based optimizer. During fitting, the categorical vehicle selection is represented by the simulator's differentiable relaxation. After optimization, it is resolved to a single candidate vehicle. The resulting descriptors $\{\mathbf{d}_i^{\star}\}_{i=1}^{n}$ are used as the input table to the DP synthesis stage.

\paragraph{Loss Function.} The loss function $\mathcal{L}$ essentially is a weighted sum of differences between the simulated and real power traces. Each trip $i$ is scored by three relative residuals between its real power trajectory $\mathbf{W}_i$ and the simulated power trajectory $\mathbf{\hat{W}}_i$: 

\begin{enumerate}
      \item \textbf{Energy Residual:} The first residual is the total energy consumed during the trip. For a given sample interval of $\Delta t = 1$\,s, the total energy is computed as the sum of the power trace over all time steps: $E_i = \Delta t \cdot \sum_t \mathbf{W}_{i,t}$.
      \item \textbf{Duration Residual:} The second residual is the total duration of the trip, which is simply the number of time steps in the power trace. However, since the duration of a trip is defined by the number of time steps in the power trajectory before the trace falls to zero, we define a soft duration residual that is computed as follows: 
      \begin{equation}
            D_i = \Delta t \sum_t \left( 1 - e^{-((p_{i,t} / \theta P)^2)} \right)
            \label{eq:softduration}
      \end{equation}
      where $p_{i,t}$ is the power at time step $t$, $\theta$ is a threshold parameter, and $P$ is the span of power to road-load model admits over its public parameter. This soft duration residual allows for a differentiable approximation of the trip duration necessary for gradient-based optimization.
      \item \textbf{Shape Residual:} The third residual is the shape of the power trace. It is computed as the cumulative distribution function (CDF) of the power trace. The shape residual is defined as follows:
      \begin{equation}
            s_{i,t} = \Big(\textstyle\sum_{u \le t} p_{i,u}\Big) \Big/ \Big(\textstyle\sum_{u \le T} p_{i,u}\Big)
            \label{eq:shape}
      \end{equation}
      where $p_{i,t}$ is the power at time step $t$ and $T$ is the total number of time steps for the power trace. The shape residual captures the distribution of power over time, which is important for accurately reconstructing the power trace.
\end{enumerate}

The loss function $\mathcal{L}$ over all $n$ trips is defined as the weighted sum of the three residuals, with weights $\lambda_E$, $\lambda_D$, and $\lambda_S$ for energy, duration, and shape, respectively. The loss function is given by
\begin{equation}
\begin{split}
  \mathcal{L} &= \frac{1}{n} \sum_{i} \Bigg[
      \lambda_E \left( \frac{\hat{E}_i - E_i}{E_i} \right)^{2}
     + \lambda_D \left( \frac{\hat{D}_i - D_i}{D_i} \right)^{2} \\
    & + \frac{\lambda_S}{T} \sum_t \big( \hat{s}_{i,t} - s_{i,t} \big)^{2}
      \Bigg]
\end{split}
  \label{eq:loss}
\end{equation}

\subsection{Data Synthesis in the Representation Space (AIM)}
\label{sec:aim}
For the synthesis in \prisma's representation space, we use the Adaptive and Iterative Mechanism (AIM)~\cite{aim}. We choose AIM since it was the best-performing mechanism in the \textsc{TabBench} evaluation~\cite{tabBench}. Here, we summarise the aspects of AIM relevant to our work.

AIM follows the \emph{select-measure-generate} paradigm common to marginal-based DP synthesis mechanisms.
Under this paradigm, algorithms iteratively select a set of queries (such as low-dimensional marginals) to measure, privately measure them by adding noise, and then post-process the noisy measurements to construct a generative model from which synthetic records are sampled.

\paragraph{Data Model and Workload Error}
Let $D$ be a sensitive dataset consisting of $N$ records, where each record $x \in D$ is a $d$-tuple over a finite domain $\Omega = \Omega_1 \times \dots \times \Omega_d$. A marginal query $M_r$ for a subset of attributes $r \subseteq [d]$ computes a vector of counts over the marginal domain $\Omega_r$, which is of size $n_r = |\Omega_r|$. A workload $W = \{(r_1, c_1), \dots, (r_k, c_k)\}$ consists of a collection of marginal queries $r_i$ paired with positive weights $c_i \ge 0$. The utility of the synthetic dataset $\hat{D}$ is evaluated using the weighted, normalized $L_1$ workload error [10]:
\begin{equation}
\text{Error}(D, \hat{D}) = \frac{1}{k \cdot |D|} \sum_{i=1}^k c_i \| M_{r_i}(D) - M_{r_i}(\hat{D}) \|_1
\end{equation}

\paragraph{Algorithmic Components}
AIM optimizes the select-measure-generate pipeline by incorporating four primary design enhancements over the earlier MWEM+PGM~\cite{pgm} instantiation of this paradigm:
\begin{enumerate}
    \item \emph{Initialization:} Rather than simply initializing with a uniform distribution, AIM allocates a small portion of the privacy budget to privately measure all 1-way marginals in the workload's downward closure. This yields a robust initial independent distribution, saving privacy budget that would otherwise be wasted selecting these fundamental attributes in early rounds.
    \item \emph{Candidates:} The candidate set $C_t$ at round $t$ is drawn from the downward closure of the workload, defined as $W^+ = \{r \mid r \subseteq s, s \in W\}$. To ensure computational tractability, candidates are filtered so that the junction tree size of the Private-PGM~\cite{pgm} engine remains below a pre-specified limit, guaranteeing a predictable memory footprint and runtime.
    \item \emph{Workload-Aware Quality Score:} Candidates are selected via the Exponential Mechanism using an improved quality score $q_r(D)$ that accounts for both expected $L_1$ error improvement and workload relevance. It is given by
    \begin{equation}
    q_r(D) = w_r \left( \| M_r(D) - M_r(p_{t-1}) \|_1 - \sqrt{\frac{2}{\pi}} \sigma_t n_r \right),
    \end{equation}
    where $w_r = \sum_{s \in W} c_s |r \cap s|$ represents the overlap weight of marginal $r$ with the workload queries. The penalty term $\sqrt{2/\pi} \sigma_t n_r$ represents the expected $L_1$ noise error, discouraging the selection of high-dimensional marginals with low signal-to-noise ratios. The sensitivity of the quality score is bounded by $\Delta_t = \max_{r \in C_t} w_r$.
    \item \emph{Adaptive Rounds and Budget:} AIM does not require specifying the number of rounds, making it hyperparameter-free. It checks if the information gain in round $t$ is small, i.e.~$\| M_{r_t}(p_t) - M_{r_t}(p_{t-1}) \|_1 \le \sqrt{2/\pi} \sigma_t n_{r_t}$. If triggered, AIM doubles $\epsilon_{t+1}$ and halves $\sigma_{t+1}$, allowing it to execute more rounds under larger privacy budgets and fewer rounds under tight budgets.
\end{enumerate}

AIM satisfies $\rho$-zero-Concentrated Differential Privacy ($\rho$-zCDP). The Gaussian mechanism satisfies $\frac{1}{2\sigma^2}$-zCDP and the Exponential mechanism satisfies $\frac{\epsilon^2}{8}$-zCDP. 
Each round consumes budget from exactly one invocation of the exponential mechanism (selection) and one invocation of the Gaussian mechanism (measurement).
By using a zCDP privacy filter~\cite{rogers_privacy_2016, feldman_individual_2021, cesar_bounding_2021}, %
AIM monitors the cumulative privacy loss $\rho_{used} = \sum_t \left( \frac{\epsilon_t^2}{8} + \frac{1}{2\sigma_t^2} \right)$ and terminates when the budget is exhausted~\cite{aim}.

\subsection{Decoding}
\label{sec:decoding}

Decoding re-runs the same forward model (Figure ~\ref{fig:forward-simulation}) on each representation -- private, for the AIM-released rows, or raw, for the non-private reference condition described in Section~\ref{sec:setup}-- using that representation's own endpoints for routing and its own aggressiveness. The only difference from encoding is that the vehicle choice is now resolved as a hard lookup rather than a soft one-hot, since the encoder (or AIM) has already committed to a single candidate. Decoding reads no real telemetry: its only inputs are the descriptor itself and public reference data (the road network, the EPA vehicle catalogue, and terrain/elevation data). Consequently, whichever DP guarantee a descriptor carries is inherited, unchanged, by the trace decoded from it, by the post-processing property of differential privacy~\cite{dwork_algorithmic_2014}.

\section{eVED Dataset}
\label{sec:eved}

For training and evaluation of \prisma, we use the eVED dataset~\cite{11174415}, which is a large-scale, diverse and public real-world dataset of vehicle energy consumption traces collected from a variety of different vehicles. It contains a collection of trips, each consisting of a time series of power consumption measurements, along with associated metadata such as vehicle type, route information, and environmental conditions.

The ground truth data is assembled from the eVED dataset by selecting two columns: (i) the power consumption time series, and (ii) the GPS coordinates of the start and end points of each trip. The power consumption time series is a sequence of measurements taken at regular intervals, representing the energy consumed by the vehicle during the trip. The GPS coordinates provide information about the route taken by the vehicle, which can be used to extract additional features such as road type, elevation changes, and traffic conditions.

\section{Experimental Setup}
\label{sec:setup}

We train and evaluate \prisma on a ground-truth data set of real trip traces, each with $500$ time steps. The ground truth data was obtained and preprocessed as described in Section~\ref{sec:eved}. We randomly split the data and fit the encoder on $10{,}000$ traces, holding out $2{,}000$ of the trips as a test set.

We run \prisma $n$ times with different random seeds for fitting the encoder. For each run, we use a privacy budget of $\varepsilon = 0.5$ for fitting the AIM synthesis, plus two non-private references:
In the first, we train AIM on the ride descriptors non-privately. 
In the second, we decode the raw representation from Section ~\ref{sec:Latent Representation} directly, without going through AIM at all. This reference provides a utility ceiling against which the effect of privatization at each $\varepsilon$ can be measured.

\subsection{Metrics}
\label{subsec:metrics}

We compare the same $2{,}000$ held-out power consumption time series against $2{,}000$ synthetic samples of each model. 
For this comparison, we used the following metrics.

\paragraph{Fréchet distance.} We fit a multivariate Gaussian (mean and full $T \times T$ covariance) to each population directly in the $T$-dimensional space of raw traces, and compute the closed-form Fréchet distance between them, normalized by the squared robust amplitude scale of the real population. %
  
  \paragraph{Autocorrelation distance.} For every time series, we calculate the empirical autocorrelation function (ACF) at lags $1,\dots,L$, $L=\min(32,,T-2)$, after centering and variance-normalizing the sequence, rendering the statistic scale-free by construction so that amplitude discrepancies are captured exclusively by the Wasserstein-based metrics rather than conflated here. Per-sequence ACF curves are averaged within each population, and we report the mean absolute deviation between the resulting real and synthetic curves, lag by lag. This isolates fidelity to the temporal dependency structure — how quickly a trace decorrelates from its own recent past — independent of its marginal amplitude distribution.
  
  \paragraph{Precision.} Following ~\cite{kynkaanniemi2019improved}, we embed both populations into a fixed-length per-sequence summary-statistic space (mean, variance, skewness, ten autocorrelation coefficients, four spectral-band energies), construct a $k$-nearest-neighbor ball around every real point using the distance to its $k$-th nearest real neighbor as radius ($k=3$), and report the fraction of synthetic points falling inside at least one such ball. Precision measures sample-level fidelity: whether individual synthetic traces are plausible members of the real manifold, independent of whether the synthetic population as a whole covers that manifold's extent.
  \paragraph{Recall.} The complementary quantity under the same construction: $k$-nearest-neighbor balls are instead built around every synthetic point, and we report the fraction of real points falling inside at least one. Recall measures diversity, i.e.\,whether the generated samples reach every mode of the real distribution. 

We additionally report FID and ACF distance as a multiple of a real vs.\,real null floor, which is computed between two random halves of the real holdout using the same metric, averaged over $15$ splits.

\subsection{Baseline Models}
\label{subsec:baselines}
\label{sec:baselines}

We compare \prisma against eight baseline models and include both non-private and DP-SGD-trained variants. For the DP variants, we retain the corresponding non-private architecture and generation procedure and differ only in their training procedure. All models are trained and evaluated on the same $10{,}000$-ride training split and $2{,}000$-ride holdout.

\paragraph{B0: Closed-form statistical release.}
B0 is a non-learning baseline that releases the mean power-consumption curve without noise and applies differential privacy only to a per-ride energy scaling factor. It clips each ride's energy to a public range $[0,E_\text{max}]$. 
We set $E_\text{max}$ to $10^8$. 
B0 then releases a DP histogram of the resulting values using the Gaussian mechanism. At generation time, it then samples an energy from the noisy histogram to rescale the mean curve. The privacy budget may be split between the histogram and, in earlier variants, a curve-shape statistic. B0 provides a deliberately simple utility floor: models that fail to outperform a mean curve with a single private scalar do not capture meaningful individual-level structure.

\paragraph{Diffusion and flow-based models.}
We evaluate four diffusion- or flow-based models that generate power trajectories through iterative denoising or numerical integration.

\begin{itemize}
    \item \emph{DDPM.} DDPM \cite{ho_denoising_2020} uses a convolutional U-Net trained to predict Gaussian noise at randomly sampled diffusion steps using an MSE objective. The generation follows the standard ancestral reverse process. The default configuration uses $1000$ diffusion steps, a linear $\beta$ schedule from $10^{-4}$ to $0.02$, a base width of $64$, depth $4$, and $200$ training epochs. 
    In contrast to Ho et al., we use a 1D rather than a 2D image U-Net. 
    We use this network as one non-private baseline. 
    \item \emph{Flow Matching.} Flow Matching uses the same 1D U-Net backbone but learns a velocity field with the rectified-flow objective of \cite{liu_flow_2023, lipman_flow_2023}. The model regresses the constant velocity $x_1-x_0$ along the interpolation $x_t=(1-t)x_0+tx_1$, with $t\sim\mathcal{U}[0,1]$. Generation solves the resulting ODE using deterministic Euler integration with $100$ steps. 
    \item \emph{DiffWave.} DiffWave \cite{kong_diffwave_2021} uses a WaveNet-style~\cite{oord_wavenet_2016} dilated-convolution architecture adapted from the original audio implementation to the $500$-step sequences considered in this work. The residual stack is reduced to $12$ layers with a dilation cycle of $8$. 
    The model predicts diffusion noise using the implementation's L1 loss and generates samples through ancestral reverse diffusion. 
    Since, in contrast to the mel-spectrogram conditioning from the original paper, a power-consumption curve has no analogous signal to condition on, we train the model in an unconditioned way. 
    \item \emph{CSDI/DiffTime.} CSDI \cite{tashiro2021csdi} is evaluated in its unconditional setting instead of its original imputation setting. Unlike for DiffWave, for this model, this modification is substantive since CSDI is originally designed for conditional imputation, not unconditional generation. Its denoising network is run with no observed-value/imputation mask conditioning at all. It is therefore close to the DiffTime~\cite{coletta_constrained_2023} implementation rather than exactly reproducing CSDI's imputation task. 
    Also, we do not use CSDI's own mask-aware training and instead rebuild the side information conditioning tensor as a plain sinusoidal time-position embedding concatenated with a learned, per-feature embedding, dropping the mask term entirely. We also changed the training objective to match the $x_0$-prediction convention the other diffusion baselines in this suite use: The original CSDI network is trained to predict the injected noise, with the loss masked to the held-out imputation positions. This adapter instead trains it to predict $x_0$ directly with an unmasked MSE loss over the full sequence. 
    This baseline model uses a dual-attention transformer with four layers, $64$ channels, and $8$ attention heads. The model predicts $x_0$ at randomly sampled steps of a $50$-step diffusion schedule, followed by ancestral reverse diffusion at generation time. 
\end{itemize}

\paragraph{Autoregressive recurrent models.} 
The autoregressive baseline uses a single recurrent GRU~\cite{cho_learning_2014} or LSTM~\cite{hochreiter_long_1997, chung_empirical_2014} cell layer and predicts a Gaussian distribution $(\mu,\log\sigma^2)$ for the next timestep. 
Training uses teacher forcing and Gaussian negative log-likelihood. The generation proceeds autoregressively from a fixed zero start token. We report GRU and LSTM as separate configurations. 
Gradient-norm clipping (default maximum norm $1.0$) is applied during training to mitigate error accumulation under autoregressive sampling. 

\paragraph{TimeGAN.}
TimeGAN \cite{yoon_time-series_2019} combines an Embedder, Recovery network, Generator, Supervisor, and Discriminator to model both the marginal distribution and temporal dynamics of sequential data. The model first learns a reversible mapping between the observation space and a lower-dimensional latent space through non-adversarial reconstruction training. 
It then pretrains the Supervisor using a stepwise supervised objective that encourages the latent representation to capture temporal transitions. Finally, the components are jointly optimized through adversarial training, combining adversarial, supervised, and moment-matching objectives. 
For the DP variant, we train the Embedder, Recovery network, Supervisor, and Discriminator using DP-SGD, as these components are trained with access to the real data.

\begin{table*}[t!]
    \centering
    \caption{Baseline comparison at $\varepsilon = 0.5$ and $\varepsilon = \infty$. For each model, the non-DP row uses no privacy mechanism at all. The $\varepsilon = 0.5$ row is that same model trained under DP-SGD. 
  All values are reported as mean $\pm$ standard deviation over $15$ seeds. FID$\times$floor is reported as a multiple of its own real-vs-real null floor, which is computed between two random halves of the real holdout using the same metric, averaged over $15$ splits.}

\centering
\begin{tabular}{lcccc}
\toprule
Model & ACF dist.$\times$floor $\downarrow$ & FID$\times$floor $\downarrow$ & Precision $\uparrow$ & Recall $\uparrow$ \\
\midrule
\multicolumn{5}{l}{\textbf{With privacy} ($\varepsilon = 0.5$)} \\
\midrule
B0            & $81.099 \pm 11.472$ & $11.252 \pm 0.971$   & ${1.000} \pm 0.000$ & $0.340 \pm 0.345$ \\
TimeGAN       & $61.240 \pm 13.152$ & $4.589 \pm 0.873$    & $0.982 \pm 0.006$ & $0.915 \pm 0.025$ \\
GRU           & ${14.746} \pm 6.782$ & $12.189 \pm 14.309$ & $0.981 \pm 0.005$ & $0.781 \pm 0.044$ \\
LSTM          & $15.470 \pm 12.588$ & $6.037 \pm 5.292$    & $0.975 \pm 0.011$ & $0.848 \pm 0.076$ \\
DiffWave      & $24.259 \pm 3.640$  & $4.527 \pm 0.152$    & $0.987 \pm 0.007$ & $0.551 \pm 0.041$ \\
DDPM          & $33.230 \pm 5.487$  & $850.675 \pm 321.032$& $0.989 \pm 0.014$ & $0.996 \pm 0.004$ \\
Flow Matching & $15.616 \pm 3.467$  & $2.426 \pm 0.212$    & $0.984 \pm 0.005$ & $0.844 \pm 0.028$ \\
CSDI          & $26.191 \pm 3.394$  & $10.282 \pm 0.301$   & $0.999 \pm 0.002$ & $0.573 \pm 0.053$ \\
\cmidrule(lr){1-5}
PRISMA + AIM       & $15.884 \pm 2.177$ & ${1.916} \pm 0.056$ & $0.999 \pm 0.001$ & ${0.999} \pm 0.001$ \\
PRISMA AE          & \multicolumn{4}{c}{\textit{no mechanism --- non-private reference only}} \\
PRISMA (untrained) & $25.271 \pm 3.848$ & $2.272 \pm 0.064$ & $0.999 \pm 0.001$ & $0.999 \pm 0.001$ \\
\midrule[\heavyrulewidth]
\multicolumn{5}{l}{\textbf{Without privacy} ($\varepsilon = \infty$)} \\
\midrule
B0            & $82.330 \pm 10.912$ & $10.087 \pm 0.136$ & ${1.000} \pm 0.000$ & $0.401 \pm 0.101$ \\
TimeGAN       & $22.193 \pm 2.150$  & $2.939 \pm 0.176$  & $0.981 \pm 0.011$ & $0.502 \pm 0.045$ \\
GRU           & $17.240 \pm 10.631$ & $6.598 \pm 8.105$  & $0.980 \pm 0.009$ & $0.850 \pm 0.211$ \\
LSTM          & $15.856 \pm 7.083$  & $5.060 \pm 3.749$  & $0.981 \pm 0.010$ & $0.855 \pm 0.148$ \\
DiffWave      & $25.966 \pm 4.829$  & $3.978 \pm 0.223$  & $0.983 \pm 0.001$ & $0.758 \pm 0.021$ \\
DDPM          & ${2.962} \pm 0.366$ & $1.334 \pm 0.979$ & $0.985 \pm 0.004$ & $0.985 \pm 0.007$ \\
Flow Matching & $6.928 \pm 1.753$   & ${0.773} \pm 0.040$ & $0.981 \pm 0.004$ & $0.982 \pm 0.001$ \\
CSDI          & $18.507 \pm 9.929$  & $2.786 \pm 1.062$  & $0.985 \pm 0.003$ & $0.980 \pm 0.008$ \\
\cmidrule(lr){1-5}
PRISMA + AIM       & $14.502 \pm 1.911$ & $1.987 \pm 0.054$ & $0.999 \pm 0.001$ & ${0.999} \pm 0.001$ \\
PRISMA AE          & $12.721 \pm 1.682$ & $1.994 \pm 0.032$ & $0.999 \pm 0.000$ & $0.999 \pm 0.001$ \\
PRISMA (untrained) & $23.559 \pm 3.104$ & $2.637 \pm 0.045$ & $0.999 \pm 0.000$ & $0.999 \pm 0.001$ \\
\bottomrule
\end{tabular}

    \label{tab:placeholder}
\end{table*}

\begin{figure*}[t!]
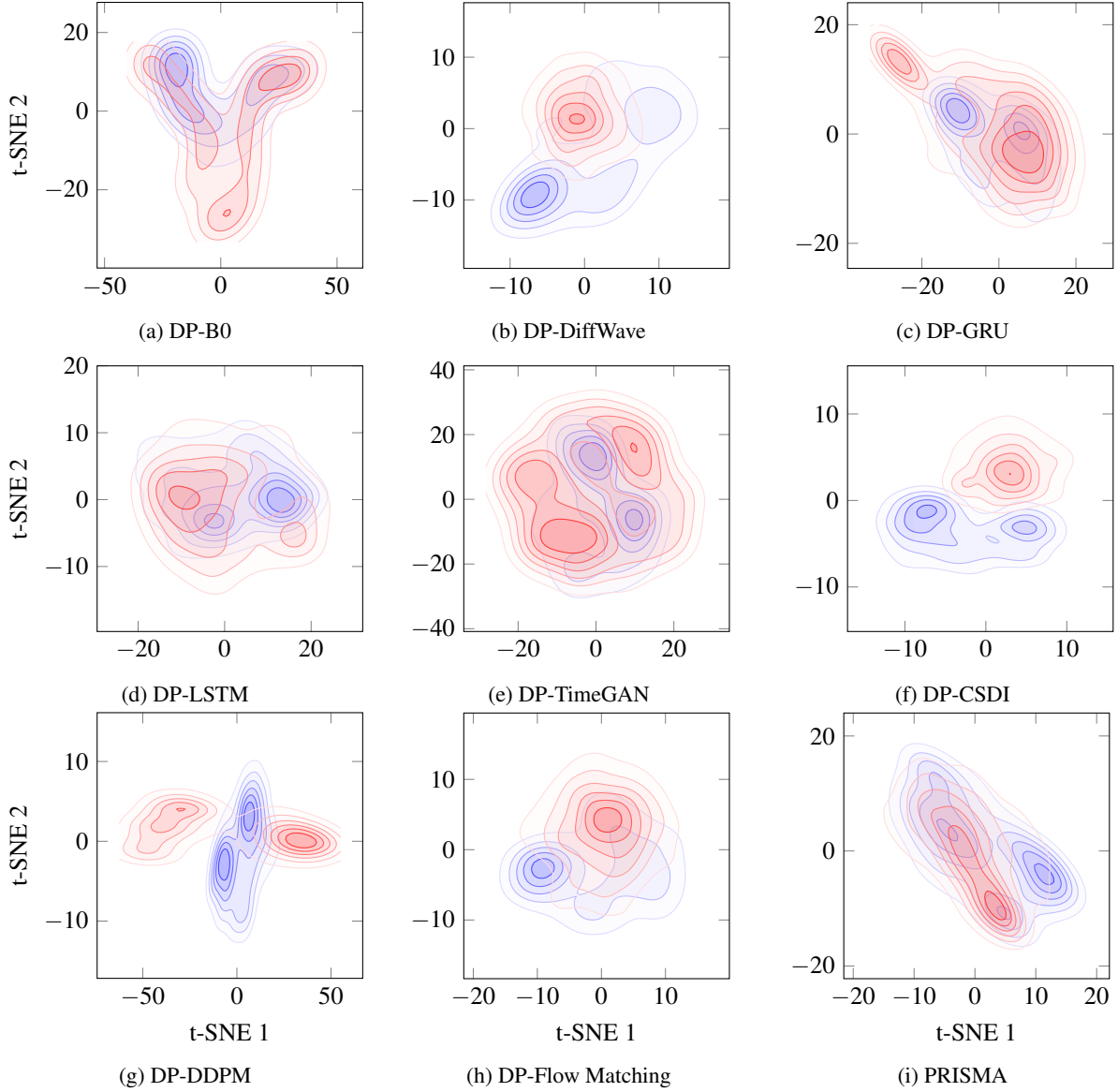

  \centering
  \begin{subfigure}{0.3\textwidth}
    \centering
    \input{images/tex/tsne/b0_contour}
    \caption{DP-B0}
  \end{subfigure}
  \begin{subfigure}{0.3\textwidth}
    \centering
    \input{images/tex/tsne/dp_diffwave_contour.tex}
    \caption{DP-DiffWave}
  \end{subfigure}
  \begin{subfigure}{0.3\textwidth}
    \centering
    \input{images/tex/tsne/dp_gru_lstm_gru_contour}
    \caption{DP-GRU}
  \end{subfigure}
  \begin{subfigure}{0.3\textwidth}
    \centering
    \input{images/tex/tsne/dp_gru_lstm_lstm_contour}
    \caption{DP-LSTM}
  \end{subfigure}
  \begin{subfigure}{0.3\textwidth}
    \centering
    \input{images/tex/tsne/dp_timegan_contour.tex}
    \caption{DP-TimeGAN}
  \end{subfigure}
  \begin{subfigure}{0.3\textwidth}
    \centering
    \input{images/tex/tsne/dp_csdi_contour.tex}
    \caption{DP-CSDI}
  \end{subfigure}
  \begin{subfigure}{0.3\textwidth}
    \centering
    \input{images/tex/tsne/dp_ddpm_contour.tex}
    \caption{DP-DDPM}
  \end{subfigure}
  \begin{subfigure}{0.3\textwidth}
    \centering
    \input{images/tex/tsne/dp_flow_matching_contour.tex}
    \caption{DP-Flow Matching}
  \end{subfigure}
  \begin{subfigure}{0.3\textwidth}
    \centering
    \input{images/tex/tsne_prisma/standard_contour}
    \caption{PRISMA}
  \end{subfigure}
  \caption{t-SNE density contour plots of \textcolor{blue}{real} (blue) vs.\ \textcolor{red}{synthetic} (red) sequences for all compared models and $\varepsilon=0.5$.}
  \label{fig:tsne_dp_models}
\end{figure*}

\section{Results and Discussion}
\label{sec:results}

All models are trained on the same $10{,}000$-ride training split and evaluated on the same $2{,}000$-ride holdout (Section~\ref{sec:setup}). Reported values are the means and standard deviations over 15 random seeds for each model and privacy budget. Both distances are normalized by the real vs.\,real null floor, which is computed between two random halves of the real holdout using the same metric. The FID and autocorrelation distance are reported as multiples of this floor, while precision and recall are reported as absolute values.

\subsection{Fidelity under Privacy}

Table~\ref{tab:placeholder} and Figure~\ref{fig:barplot} summarize both conditions. The non-private rows show that the diffusion and flow-based models are the strongest models in the comparison, with \textsc{Flow Matching} and \textsc{DDPM} achieving the best FID and autocorrelation distance. The autoregressive models are mid-field, while \prisma is slightly worse than the best baseline but still competitive. The non-learning baseline B0 performs poorly, as expected. Under $\varepsilon = 0.5$, the diffusion and flow-based models degrade substantially, while \prisma's fidelity is largely unchanged. \textsc{Flow Matching} degrades by a factor of $3.1\times$ in FID, \textsc{DDPM} by $638\times$, while \prisma's FID is unchanged within noise ($0.96\times$). This makes it the most faithful model in the comparison.

The precision and recall results largely reflect the models' constructions rather than differences in training. Precision is nearly saturated for every model in both conditions ($\geq 0.975$). Although \textsc{B0} achieves the nominal maximum of $1.000$, it produces only a single rescaled mean curve. Its non-private recall of $0.40$ reveals the resulting lack of coverage. Recall is therefore more informative here. \prisma achieves the highest recall in both conditions and remains unchanged under privacy ($0.999$). In contrast, \textsc{Flow Matching}, \textsc{CSDI}, and \textsc{DiffWave} lose coverage under privacy, reaching $0.844$, $0.573$, and $0.551$, respectively. \textsc{DDPM} is an instructive exception: its recall increases to $0.996$, while its FID worsens by $638\times$, indicating increased diversity at the expense of fidelity.

This pattern reflects a fundamental difference in how the models interact with the privacy mechanism. Because \prisma operates in a low-dimensional space, the noise it must absorb is small and its fidelity is largely unaffected. The deep generative baselines, by contrast, privatize high-dimensional raw telemetry, and the much larger noise budget required renders their outputs unfaithful. In other words, the physical representation is not what makes \prisma accurate in the non-private setting--it is what makes it resilient when privacy is enforced.

\subsection{Ablation Studies}

Table~\ref{tab:placeholder} reports two ablations: (i) applying the encoder and decoder without AIM (PRISMA AE) as a non-private upper bound, and (ii) AIM applied to an unfitted representation (PRISMA untrained) as a lower bound on fidelity.

\paragraph{Ablation: no mechanism.} This ablation uses the encoder and decoder, removing AIM and providing a non-private upper bound on fidelity. On ACF distance, this ablation behaves as a ceiling at $12.72\times$ the real vs.\,real floor against $15.88\times$ for the full pipeline. This suggests that the encoder and decoder effectively capture the underlying structure of the data, and the mechanism and noise together are responsible for roughly a fifth of the degradation in fidelity. On FID, the ablation is at $1.99\times$ the real vs.\,real floor against $1.92\times$ for the full \prisma pipeline. Thus, the encoder and decoder account for most of the fidelity loss, while AIM has only little measurable impact. 

\paragraph{Ablation: no fitting.} This ablation uses AIM applied to a representation that was never fitted, which provides a lower bound on fidelity. Its autocorrelation distance increases to $25.27\times$ the floor and its FID to $2.27\times$, compared with $15.88\times$ and $1.92\times$ for \prisma, respectively. These results show that fitting the representation is essential for preserving fidelity.

\subsection{Distributional Structure} 
\label{sec:distributional_structure}

\begin{figure}[htb!]
    \centering
    \includegraphics[width=\linewidth]{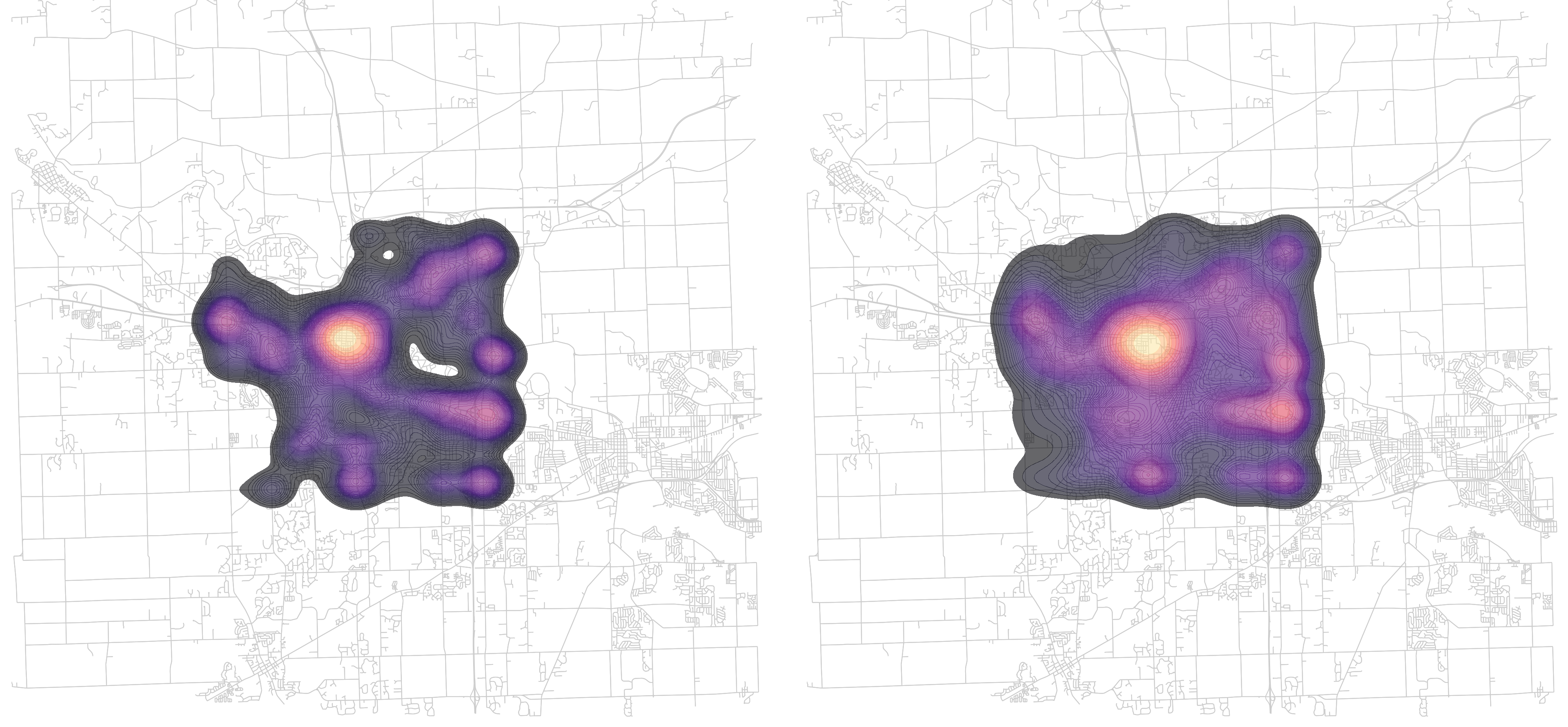}
    \caption{heatmap of the start and end coordinates of the ground-truth dataset (left) and that of the coordinates synthesized using AIM (right) with $\varepsilon=0.5$.}
    \label{fig:placeholder}
\end{figure}

Table~\ref{tab:placeholder} quantifies fidelity but does not capture distributional structure. We therefore use t-SNE to project real and synthetic sequences into two dimensions and visualize their density contours.

\paragraph{The decoded traces.} Figure \ref{fig:tsne_dp_models} shows t-SNE density contour plots of real vs.\,synthetic sequences for all mechanisms 
for $\varepsilon=0.5$. 
The plots show that \prisma is able to preserve the structure of the data better than most other models under privacy constraints.

\paragraph{The released representation.}Figure \ref{fig:placeholder} shows a heatmap of the start and end coordinates of the ground-truth dataset and the coordinates synthesized using AIM with $\varepsilon=0.5$. The heatmap indicates that \prisma is able to preserve the spatial distribution of the data even under privacy constraints.

Both the t-SNE plots and the heatmap suggest that \prisma is able to generate synthetic data that is more faithful to the underlying structure of the real data, even under privacy constraints. This is consistent with the quantitative results in Table~\ref{tab:placeholder}, which show that \prisma achieves better fidelity than the other models under privacy constraints. 

\subsection{Limitations}
\label{sec:discussion}

\paragraph{Low-dimensional representation.}
\prisma represents each trip using only three fitted quantities (aggressiveness, detour, and vehicle) beyond its endpoints. This enables privatisation in a low-dimensional space rather than over raw telemetry, but necessarily discards variation not captured by these parameters. For example, weather, auxiliary loads, and fine-grained driving behavior cannot be recovered by the decoder, irrespective of the privacy budget.

\paragraph{Dependence on public reference data.}
Reconstruction fidelity is constrained by the coverage and quality of the public road network, elevation data, and the EPA vehicle catalog used by the decoder. Trips involving vehicles poorly represented in the candidate pool, or routes that differ substantially from those selected by the public router, may therefore be difficult to reconstruct faithfully even when the underlying descriptor is accurately estimated.

\paragraph{Dataset scope.}
Our evaluation is conducted on a dataset with a specific geographic and temporal scope (Section~\ref{sec:setup}). This may limit the generality of our findings, particularly where road topology, vehicle composition, driving behavior, or environmental conditions differ compared to the eVED dataset.%

\section{Conclusion}
\label{sec:conclusion}

We presented \prisma, a differentially private synthesis pipeline for in-car energy telemetry that exploits the low intrinsic dimensionality of trip-level energy consumption. \prisma fits each trip with a compact, physically interpretable descriptor, applies the AIM mechanism only to this low-dimensional representation, and reconstructs telemetry from the resulting synthetic descriptors through privacy-free post-processing. This design avoids applying differential privacy directly to high-dimensional time series and thereby mitigates the associated dimensionality penalty. Our experimental results show that at a strong privacy budget of $\varepsilon=0.5$, \prisma retains the best fidelity of any model evaluated, and that its fidelity is essentially unchanged from the non-private setting. In contrast, the fidelity of all other models degrades substantially under privacy, with one (DDPM) losing more than $600\times$ in Fréchet distance. These results demonstrate that \prisma is a promising approach for generating high-fidelity synthetic energy telemetry under strong privacy constraints.

\section*{Ethical Considerations}
\prisma is designed to increase privacy protection by enabling the release of synthetic instead of real data. The framework provides formal $(\varepsilon,\delta)$-differential privacy guarantees that bound the privacy loss for any individual driver whose data contributes to the released statistics. 

Our empirical evaluation uses the Extended Vehicle Energy Dataset (eVED)~\cite{11174415}, a publicly available, anonymized research dataset of real-world vehicle energy consumption traces. The data was collected by the original authors in compliance with relevant data-sharing guidelines and contains no direct personal identifiers (such as driver names, vehicle identification numbers, or user IDs). Because our study involves exclusively secondary analysis of an existing public dataset without interaction with human participants, it does not constitute human-subjects research requiring Institutional Review Board (IRB) approval.

\section*{Open Science}

The source code of \prisma, including the differentiable driver and car simulations, privacy mechanism, and optimization loop, will be made publicly available upon acceptance at \url{https://anonymous.4open.science/r/PRISMA-6C17/README.md}.

\bibliographystyle{plainurl}
\bibliography{refs}

\appendix
\end{document}